%% file: Beitz_Medea_v05.tex
\documentclass[preprint,5p,times,authoryear,twocolumn]{elsarticle}

\usepackage{graphicx}
\usepackage{hyperref}
\usepackage[english]{babel}
\usepackage[ansinew]{inputenc}

\def\cms{$\mathrm{cm\,s^{-1}}$}

\def\ms{$\mathrm{m\,s^{-1}}$}
\def\mum{$\mu$m}

\journal{Icarus}

\begin{document}

\begin{frontmatter}

\title{Free Collisions in a  Microgravity Many-Particle Experiment. II.\\ The Collision Dynamics of Dust-Coated Chondrules}
\author{E. Beitz$\rm^{a}$}\ead{e.beitz@tu-bs.de}
\author{C. G\"uttler$\rm^{a,b}$}
\author{R. Weidling$\rm^{a}$}
\author{J. Blum$\rm^{a}$}
\address{$\rm^{a}$Institut f\"ur Geophysik und extraterrestrische Physik, Technische Universit\"ut Braunschweig,\\Mendelssohnstr. 3, D-38106 Braunschweig, Germany}
\address{$\rm^{b}$Department of Earth and Planetary Sciences, Kobe University, 1-1 Rokkodai-cho, Nada-ku, Kobe 657-8501, Japan}

\begin{abstract}
The formation of planetesimals in the early Solar System is hardly understood, and in particular the growth of dust aggregates above millimeter sizes has recently turned out to be a difficult task in our understanding [Zsom et al. 2010, A\&A, 513, A57]. Laboratory experiments have shown that dust aggregates of these sizes stick to one another only at unreasonably low velocities. However, in the protoplanetary disk, millimeter-sized particles are known to have been ubiquitous. One can find relics of them in the form of solid chondrules as the main constituent of chondrites. Most of these chondrules were found to feature a fine-grained rim, which is hypothesized to have formed from accreting dust grains in the solar nebula. To study the influence of these dust-coated chondrules on the formation of chondrites and possibly planetesimals, we conducted collision experiments between millimeter-sized, dust-coated chondrule analogs at velocities of a few \cms. For 2 and 3 mm diameter chondrule analogs covered by dusty rims of a volume filling factor of 0.18 and 0.35-0.58, we found sticking velocities of a few \cms. This velocity is higher than the sticking velocity of dust aggregates of the same size. We therefore conclude that chondrules may be an important step towards a deeper understanding of the collisional growth of larger bodies. Moreover, we analyzed the collision behavior in an ensemble of dust aggregates and non-coated chondrule analogs. While neither the dust aggregates nor the solid chondrule analogs show sticking in collisions among their species, we found an enhanced sicking efficiency in collisions between the two constituents, which leads us to the conjecture that chondrules might act as ``catalyzers'' for the growth of larger bodies in the young Solar System.

\end{abstract}


\end{frontmatter}

\section{Introduction}

Chondrules are mm-sized solid particles, which are found in large amounts in meteorites. Petrographic studies of these chondrules revealed that their formation dates back to an early time interval in the solar nebula, only $2.5 \pm 1.2$ Myrs after the first condensates and the calcium-aluminum-rich-inclusions formed \citep{AmelinEtal:2002}. As chondrules make up between 20 and 80 \% of their parent bodies' volume \citep{HewinsEtal:2005}, their mere ubiquity indicates a tight connection to the formation of planetesimals. \citet{MetzlerEtal:1992} analyzed fourteen carbonaceous chondrites and determined that all chondrules are surrounded by a dusty layer, referred to as accretionary rims, which are hypothesized to be a record of the accretion of dust grains in the solar nebula \citep{MetzlerEtal:1992, MorfillEtal:1998, BlandEtal:2011}. These dusty rims will likely have an influence on the collision behavior of the chondrules and, thus, on the formation of planetesimals \citep{OrmelEtal:2008}. Especially since \citet{ZsomEtal:2010} discovered a bouncing barrier for millimeter-sized dust aggregates, which prevents any further mass increase of the dust aggregates, the growth mechanisms beyond this size need further investigation. For the chondrules we firmly know that they existed before the planetesimals were formed. On the account that the formation process of the chondrules is still under discussion, it could also be likely that not all dust aggregates were consumed to form igneous particles so that the remaining dust aggregates and the freshly-formed chondrules could have co-existed.

With this in mind, we experimentally tested the influence of porous dust rims around artificial chondrules and the co-existence of chondrule analogs and dust aggregates on their collision behavior in free particle-particle collisions. To achieve this, we first developed two different techniques to cover chondrule analogs with dusty layers, simulating their formation in the solar nebula (Sect. \ref{sec:setup_coating_experiment} and \ref{sec:results_coating_experiment}). In a second step, the dust-coated chondrule analogs were used in a multiple collision experiment, in which we observed low-velocity collisions (typically at a few \cms) between dust-coated particles of 2 and 3 mm in diameter (Sect. \ref{sec:setup_collision_experiment} and \ref{sec:results_collision_experiment}). Additionally, we performed a collision experiment between dust aggregates and chondrule analogs with a diameter of 2 mm each, to test the influence of the presence of solid particles on the formation of larger bodies.

\section{Experimental Setup}

\subsection{Coating Experiments} \label{sec:setup_coating_experiment}

We used two different experimental methods to prepare dust rims around chondrule analogs, resulting in simulated accretionary rims of different porosities and morphologies. As chondrule analogs, we used glass beads of 2 and 3 mm diameter, respectively.

The first setup is designed to create highly porous dust rims in collisions between a chondrule analog and individual \mum-sized dust grains at velocities below the sticking threshold. This experiment is intended to simulate free-floating chondrules in the solar nebula, which gently collide with single dust grains, which then stick to their surfaces \citep{MorfillEtal:1998}. To reproduce such a scenario, we levitated a glass bead (the chondrule analog) in a vertical upward gas stream within a hollow funnel. The gas flow from below the chondrule analog is adjustable so that we can levitate glass spheres between 1 and 4 mm in diameter without touching the funnel wall. Using an aerosol generator, we introduced single micrometer-sized dust grains into the gas flow, which stick to the chondrule's surface. A typical dust-coated chondrule analog formed with this method is shown in the inset of Fig. \ref{fig:sicking}a and the properties of the dust rims are presented in Sect. \ref{sec:results_coating_experiment}. For this method, we determined the collision velocity between the levitated glass bead and the dust grains, assuming perfect coupling of the dust particles to the gas flow, to be $v_{\mathrm{coll}} \leq 0.8 $ \ms. This velocity determination is based on the measurement of the gap width between the levitated glass sphere and the funnel wall with a camera mounted such that the experiment could be observed from the top. Additionally, the gas flux through the funnel was measured so that the equation of continuity yields the gas velocity in the gap. This value overestimates the gas velocity and denotes the upper limit of the collision velocity of the individual dust particles with the surface, because (i) the dust particles never completely couple to the gas flow and (ii) the gas velocity at the surface of the glass bead possesses only a tangential velocity component whereas the collision velocity between dust and glass bead is dominated by the normal component. Nevertheless, the estimated upper limit is still sufficiently below the sticking velocity of 1.1 \ms\ determined by \citet{BlumEtal:2006} to guarantee a high sticking probability. Moreover, for these low velocities, we do not expect any severe restructuring \citep{BlumWurm:2000,BlumEtal:2006} or even erosion \citep{SchraeplerBlum:preprint} of the dusty surface, and this is consistent with the low volume filling factor of $\phi = 0.18\pm 0.05$ measured for the accreted dust rims (see Sect. \ref{sec:results_coating_experiment}). For the determination of the volume filling factor, we used a rotation table and a high-resolution camera (for the volume determination, resolution 4 $\rm\mu m/$pixel) and a precision balance (for the mass determination) with an accuracy of 0.1 mg, details are described by \citet[hereafter Paper I]{WeidlingEtal:preprint}. The coating is rather homogeneous over the surface of the glass bead, but due to the handling processes using tweezers and due to the storage, a small fraction of the dust-coated surface exhibits depressions or abrasions (see inset of Fig. \ref{fig:sicking}a). We consider the influence of these imperfections as minor.

The second method simulates a multiple-collision scenario in which mass is transferred to the chondrule surfaces in many low-velocity collisions with porous dust agglomerates. To assemble these rims, we randomly shake a container of dust with a glass bead inside for a defined time (e.g., 60 s). With this method, the glass beads accrete a denser rim in comparison to the first method described above. We estimated the volume filling factor to be $0.35 \leq \phi \leq 0.58$ (see Sect. \ref{sec:results_coating_experiment}). The inset in Fig. \ref{fig:sicking}b shows such a chondrule analog, which has a more irregular rim than the one shown in Fig. \ref{fig:sicking}a.

For both coating scenarios, we used spherical monodisperse SiO$_2$ dust grains with a grain diameter of 1.5 \mum\ \citep[properties compiled by][]{BlumEtal:2006}. This dust has been used for many dust-aggregate collision experiments \citep{BlumWurm:2008, GuettlerEtal:2010}, and also its material and aggregate properties are well known \citep{HeimEtal:1999, BlumEtal:2006}.

For the experiment in which we observed collisions between un-coated glass beads and dust aggregates, the dust aggregates consisted of polydisperse $\rm SiO_2$ dust grains and had a diameter of about 2 mm. These particles are the same as described in Paper I, with a volume filling factor of $\phi=0.35\pm 0.05$.

\subsection{Multiple Collision Experiments}\label{sec:setup_collision_experiment}

The main objective of this study is to investigate low-velocity collisions between the protoplanetary dust analogs (dust aggregates, pure chondrules, dust-coated chondrules) and to determine their sticking threshold velocity. To achieve the required low velocities, the experiments were performed under microgravity conditions in the Bremen drop tower, and the setup used here is the many-particle collision experiment MEDEA-II as described in detail in Paper I. The setup consists of a glass tube under vacuum (residual gas pressure $\sim$ 0.1 Pa) with a diameter of 25 mm, in which the particles can collide freely. The experiment chamber is observed by a high-speed camera operated at 500 frames per second, using a beam splitter optics with a viewing angle of $30^{\circ}$ to obtain the three-dimensional trajectory information. In this paper we use only one projection for the determination of the collision velocity. However, Paper I showed that the major component to the collision velocity is in the direction parallel to the shaking direction and that a two dimensional treatment underestimates the collision velocity by only 13 $\%$. In the case of hypothetical sticking, we used the second camera projection to ensure that sticking really occurred. Within the experimental campaign in November 2010, we performed four experiments with dust-coated chondrule analogs (using 2 and 3 mm diameter glass beads and the two rim porosities described above, we conducted one experiment for each combination), and one experiment with a mixture of solid glass beads and dust aggregates. During approx. 9 s of microgravity time, the vacuum tube was gently shaken with an eccentric wheel (1 mm amplitude, 4 or 16 Hz maximum shaking frequency) to separate the chondrule analogs and dust aggregates at the beginning of the experiment and to prevent their sticking to the walls at a later stage. In each of the four experiments with dust-coated chondrule analogs described below, 60 -- 140 particles were carefully prepared and stored at the bottom of the test chamber in a conical pile. In the experiments with the fluffier rims, the vacuum tube was continuously excited by a fast shaker (16 Hz), in the runs with the denser rims, the shaker was running at a lower frequency (4 Hz), which was even slowed down (approx. 2 Hz) after 3 s with a motor control. By this, we intended to achieve even lower collision velocities, but this had the side effect that the particles were clumping rapidly and no further individual collisions could be observed. In the fifth experiment, we used 40 un-coated glass beads and 95 dust aggregates with a diameter of 2 mm each. These are the same dust aggregates which were used in the experiments of Paper I. The dust aggregates possessed a volume filling factor of $\phi= 0.35 \pm 0.05$. The shaker frequency in this experiment was 16 Hz for the first 2 s and then reduced to 8 Hz for the remaining 7 s.

\section{Results}

\subsection{Rim Properties} \label{sec:results_coating_experiment}

The two different coating mechanisms described in Sect. \ref{sec:setup_coating_experiment} result in two different types of chondrule rims. The rims formed by free-floating chondrules in the funnel could well be characterized with the volume measurement on a rotation table as described in Paper I. For these rims, we found a typical volume filling factor of $\phi = 0.18 \pm 0.05$ (standard deviation), which is only slightly above the BPCA (Ballistic Particle-Cluster Agglomeration) limit \citep[see, e.g.,][]{BlumSchraepler:2004} and not depending on coating time. We tested this by taking the particles out of the funnel after 30, 60, and 120 s, respectively, and precisely measuring the mass and volume. The mean rim thickness increased linearly in time at a rate of 0.6 $\mathrm{\mu m\, s^{-1}}$. For the collision experiments described below, we exposed the glass beads to the dust flow for 60 to 90 s, which leads to rim thicknesses of around 50 \mum.

The second coating method is less defined, and a measurement of the rim volume did not result in plausible values due to the very uneven surface (cf. inset in Fig. \ref{fig:sicking}b). The rotation-table method systematically results in an underestimation of the filling factor and the best guess we can make for the density of the dust rims is to relate their filling factor to the aggregates in the container, which have a volume filling factor of $\phi=0.35$ (Paper I). The maximal value of $\phi=0.58$ was reached in the omnidirectional compression experiments by \citet{GuettlerEtal:2009}, which is close to random close packing, the maximum compaction possible. However, our experimental technique is likely to be more comparable to unidirectional compression \citep{BlumSchraepler:2004} that leads to volume filling factors of $\phi=0.33\pm0.02$. Therefore, we expect the volume filling factor of the accreted rim to be on the lower end of the given range.

\subsection{Collision Behavior} \label{sec:results_collision_experiment}

\begin{figure}[t]
    \begin{center}
        \includegraphics[width=8 cm]{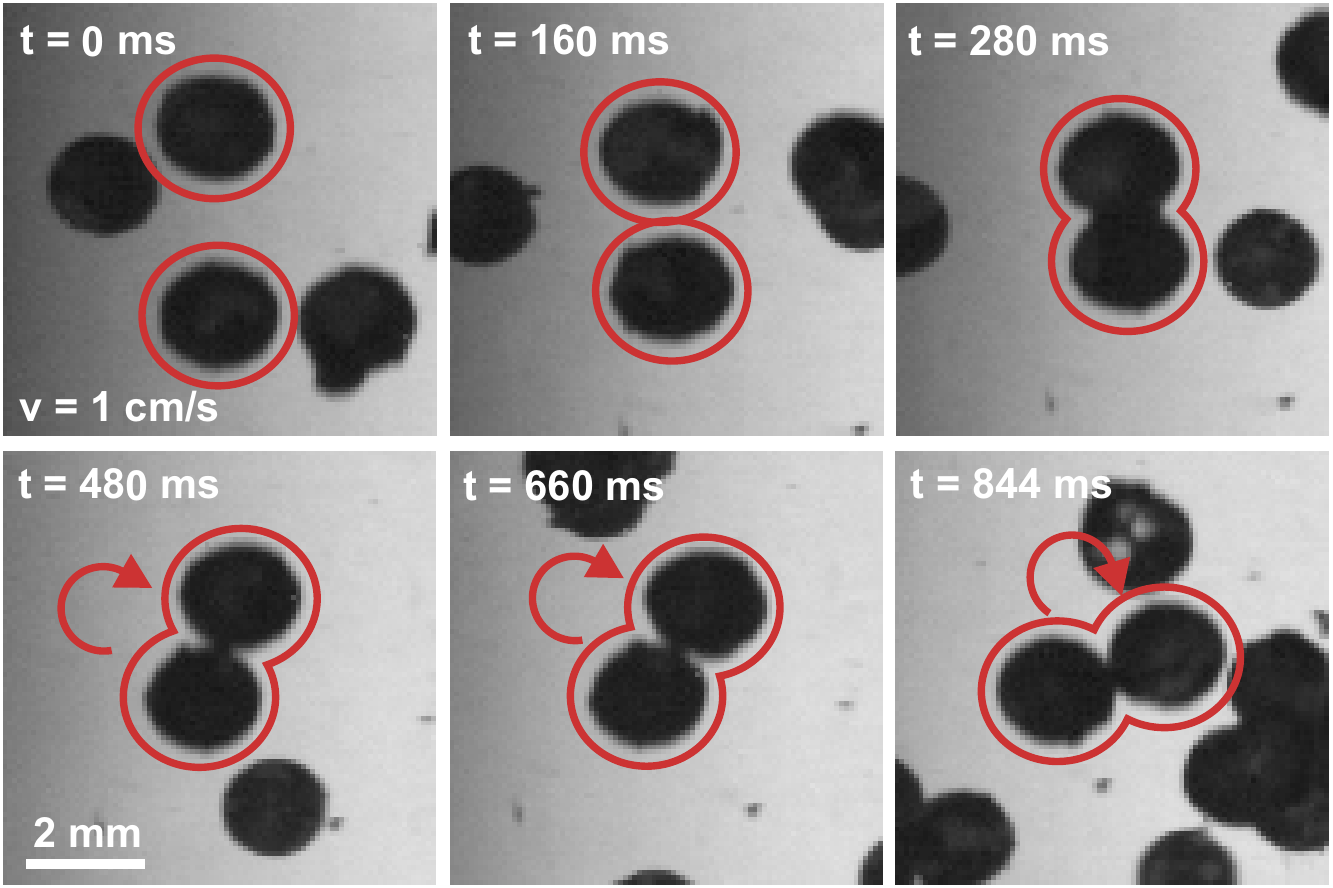}
        \caption{\label{fig:sequence}Image sequence of a sticking collision between two dust-coated chondrule analogs (2 mm diameter, $\phi= 0.18$) at a velocity of 1 \cms. After sticking to each other, they rotate around their common center of mass. A movie of this collision can be found in the online version of this article.}
    \end{center}
\end{figure}

\begin{figure*}[t]
    \begin{center}
        \includegraphics[width=\textwidth]{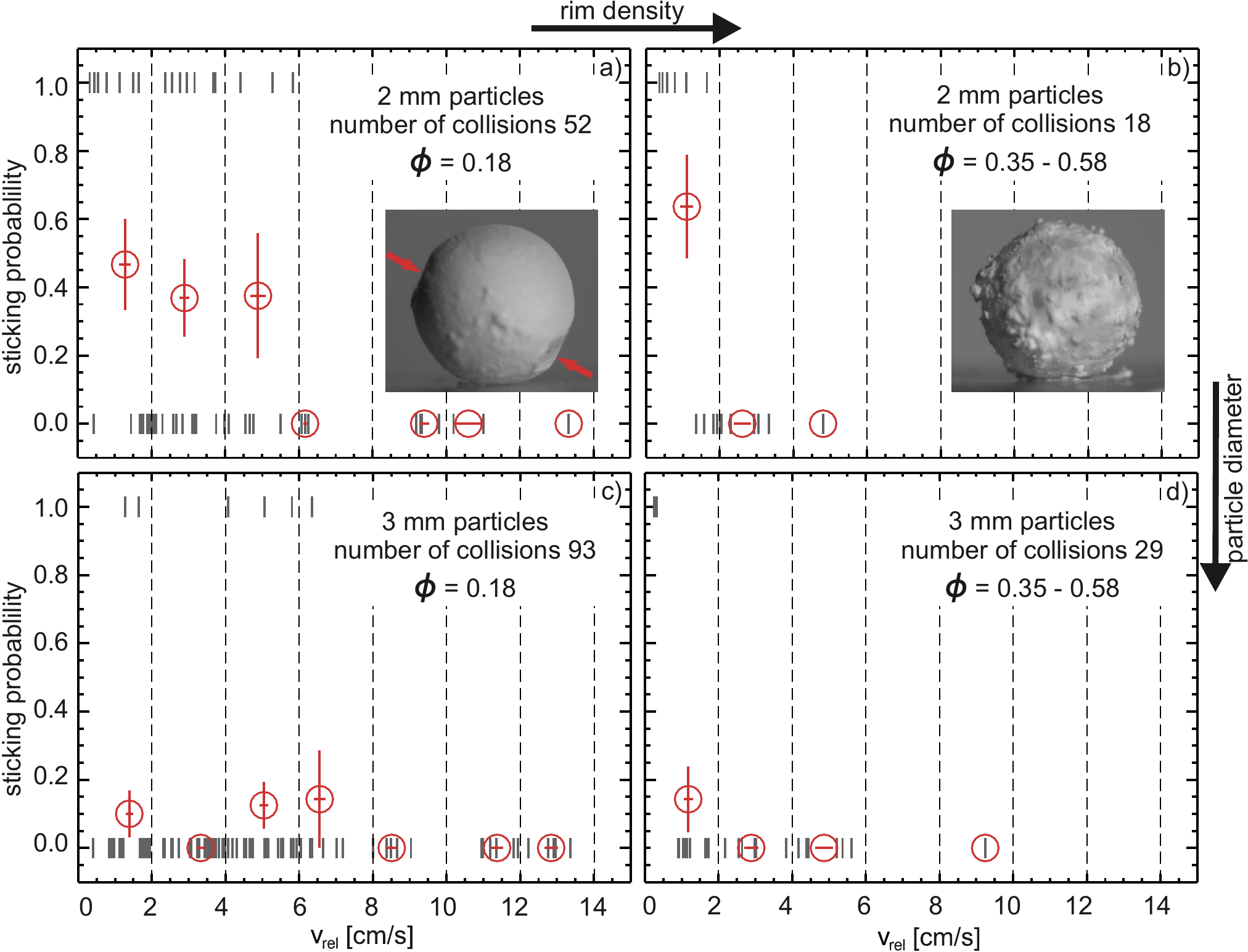}
        \caption{\label{fig:sicking}Sticking probabilities of the four experiments with two different diameters for the chondrule analog and two different rim porosities. The averaged sticking probability (circles) is a decreasing function of velocity and slightly decreases with increasing volume filling factor (to the right) and increasing particle size (downwards). The insets show examples of the dust-coated particles with different porosities produced by the two setups described in Sect. \ref{sec:setup_coating_experiment}. The arrows in a) mark the depressions due to the handling procedure.}
    \end{center}
\end{figure*}

\begin{figure}[t]
    \begin{center}
        \includegraphics[width=8 cm]{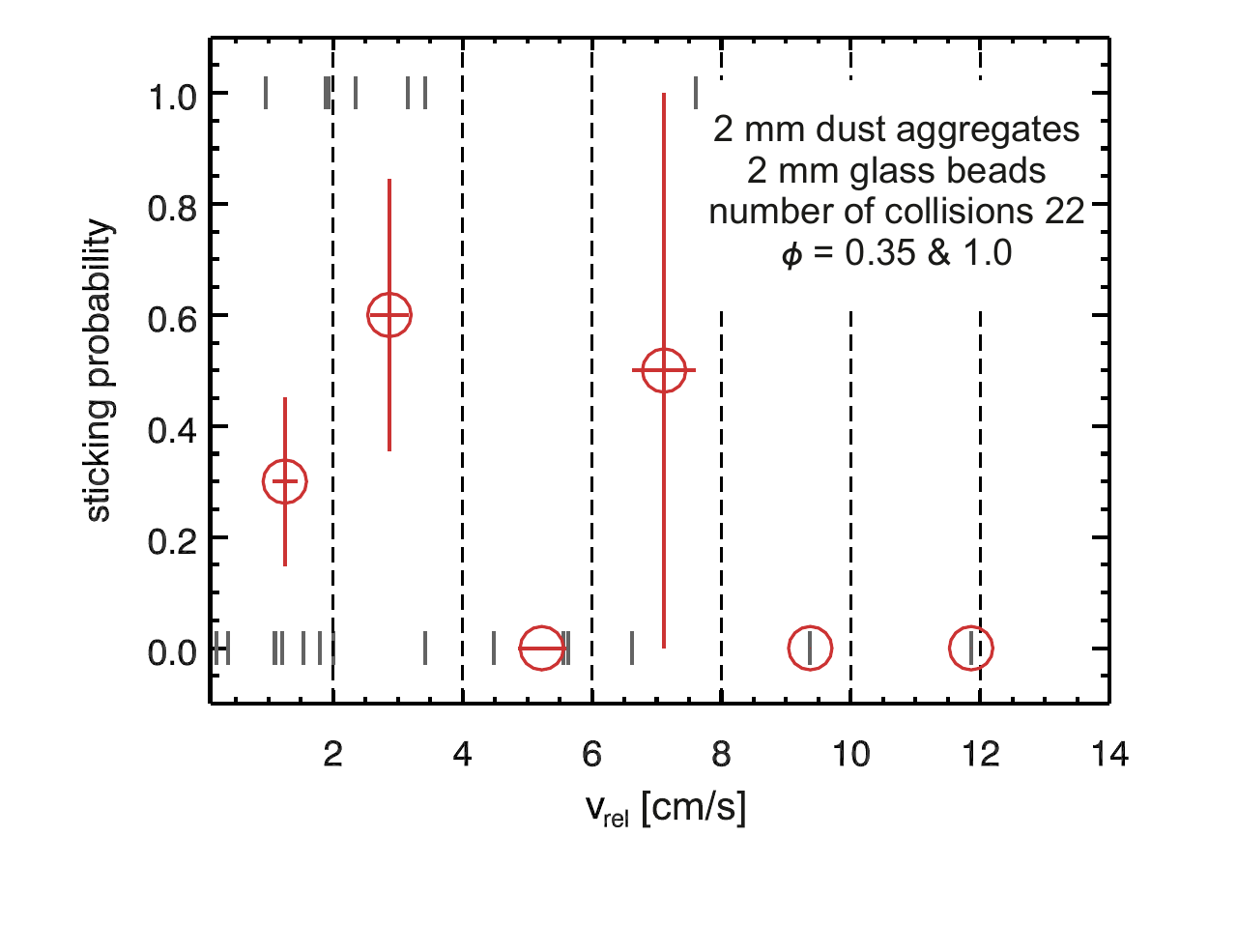}
        \caption{\label{fig:sicking_beadsanddust}Sticking probabilities of the experiment with 40 glass beads and 95 dust aggregates of the same size. The circles denote the mean value for each bin. Here, only the collisions between dust aggregates and glass beads are taken into account. Collisions between two dust aggregates as well as between two glass beads only result in bouncing.}
    \end{center}
\end{figure}

In this section, we present the results on collisions between individual 2 mm and 3 mm coated chondrule analogs as well as the results of collisions of dust aggregates with solid glass spheres. A typical sticking collision of dust-coated glass beads is presented in Fig. \ref{fig:sequence}: the two particles collide at a relative velocity of 1 \cms, stick to each other and rotate around their common center of mass. In this sequence, the dust envelope has a volume filling factor of $0.18\pm 0.05$. A movie of this collision, as well as a movie of the whole sequence in which this collision and the collision mentioned in Sec. \ref{sec:cluster_formation} are highlighted is available as online material. In total, we observed 214 individual collisions in the five experiments for which we were able to follow both particle trajectories. For those collisions, we measured the projected collision velocities and we also discerned the result into sticking or bouncing. Figure \ref{fig:sicking} shows the results of the four experiments with dust-coated chondrule analogs (two particle sizes, two porosities) in terms of the sticking probability, and Fig. \ref{fig:sicking_beadsanddust} shows the sticking probability for the experiment in which un-coated glass beads and dust aggregates collided. In this latter case, only the collisions between dust aggregates and solid spheres were analyzed, a sticking collision among the dust aggregates or between the solid glass beads could not be found. In Figs. \ref{fig:sicking} and \ref{fig:sicking_beadsanddust}, the vertical tick marks at sticking probabilities 0 or 1 denote individual bouncing or sticking collisions, respectively. We averaged the raw data in intervals of 2 \cms\ (within the boundaries denoted in Figs. \ref{fig:sicking} and \ref{fig:sicking_beadsanddust} by the dashed vertical lines), which is shown by the circles. The error bars denote the standard deviation in sticking probability and collision velocity for these mean values. For the dust-coated particles, we find that (i) the sticking probability increases with decreasing velocity, (ii) the sticking probability is higher for the smaller particles (upper row) as compared to the larger chondrule analogs, and (iii) particles stick at higher velocities if the rims are fluffier (compare left with right column). We have to note that the statistics (i.e., the total number of observed collisions) for the experiments with denser rims is much poorer than for the fluffier envelopes, because the excitation of our shaker was turned down as mentioned above. The same trend is also obvious in Fig. \ref{fig:sicking_beadsanddust} for the collisions of dust aggregates with glass beads. Moreover, as the velocities plotted here are projected velocities, they are systematically underestimated by approx. 13 \% in average (see discussion in Paper I).

\subsection{Cluster Formation}\label{sec:cluster_formation}

In the beginning of each experiment, the dust-rimmed particles as well as the dust aggregates and the glass beads had been piled at the bottom of the experiment chamber and were then levitated under microgravity conditions due to the excitation from the shaker. As the particles are extremely cohesive (see the previous section), they were initially not fully separated from each other but several clusters repeatedly collided with the vibrating top and bottom of the vacuum chamber. After a few seconds, this led to a quasi-equilibrium between growth and destruction of the clusters, except for the case were the shakers were turned off (see Sect. \ref{sec:setup_collision_experiment}).

\begin{figure}[t]
    \begin{center}
        \includegraphics[width=8 cm]{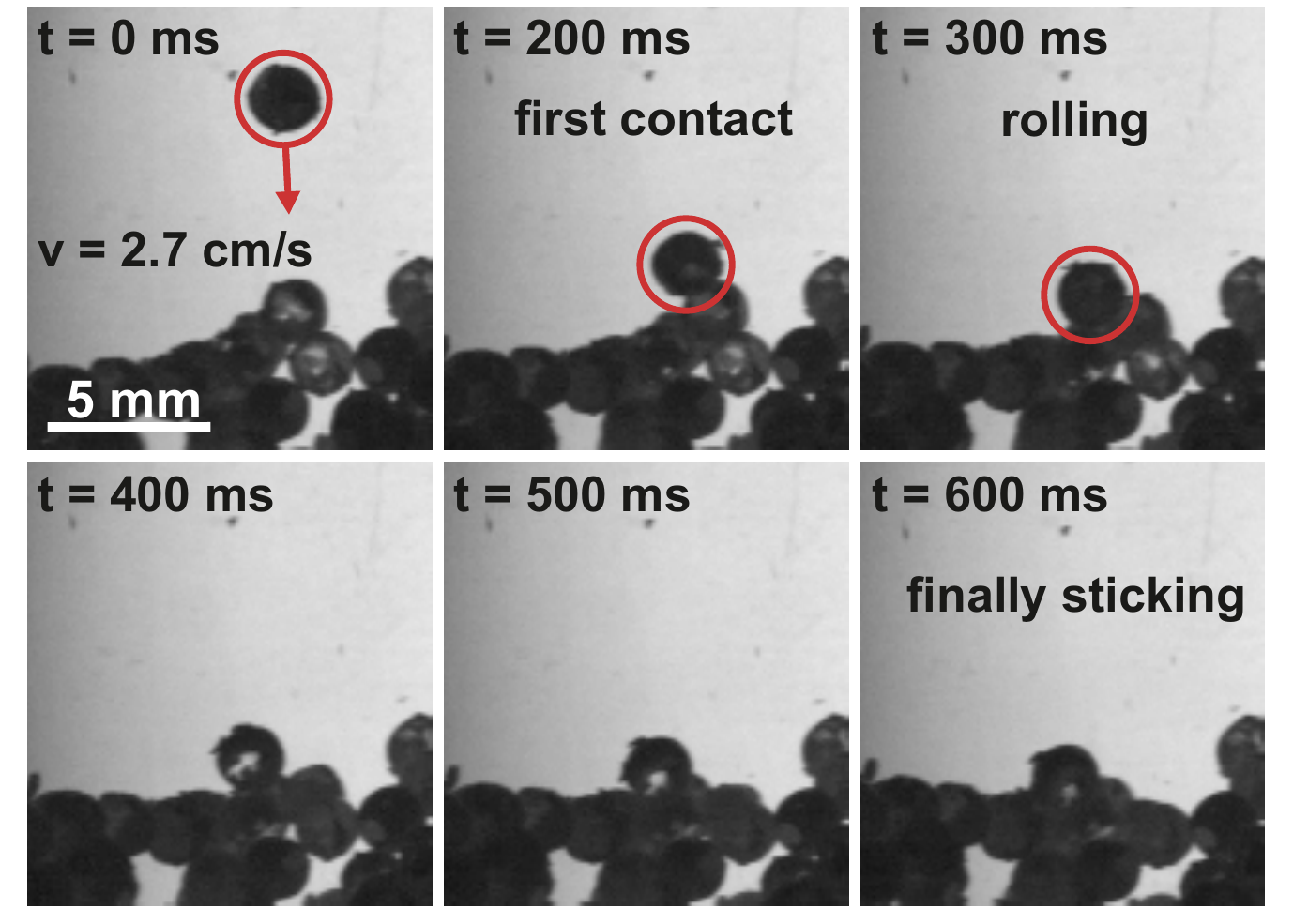}
        \caption{\label{fig:cluster_sequence}Clusters of dust-rimmed chondrules formed in each of the four experiments and were observed to be growing. Here, we show a collision of a single dust-coated chondrule analog with a chondrule-analog cluster at a velocity of 2.7 \cms, which, after some reorganization within the cluster (notable by the deformation of the cluster at the impact point), leads to sticking. A movie of this collision can be found in the online version of this article.}
    \end{center}
\end{figure}

Although we did not yet succeed in a broad quantitative analysis of the collisions among clusters and between clusters and individual particles, we qualitatively observed an efficient growth of these clusters. In both experiments with compact rims ($\phi=0.35-0.58$) we observed several collisions between individual dust-coated glass beads (as presented above) but after the shaker was slowed down, most individual particles were efficiently swept up by one central cluster. Figure \ref{fig:cluster_sequence} shows an individual 2 mm-sized particle, which is colliding with a chondrule-analog cluster at a velocity of 2.7 \cms\ (movie is available as online material). One can qualitatively see that the sticking efficiency is enhanced compared to a collision between individual particles, because the rimmed particle collides with a salient particle from the cluster, whereupon the cluster is being deformed at this position. This can be seen by following the two vertically aligned, semi-transparent particles in the line of the impact velocity over the whole sequence. The cluster deformation naturally goes along with a dissipation of energy, by which the impactor is slowed down. The impacting particle does not bounce back, but is still moving in the direction towards the cluster, in which it can only roll over the cluster surface. Thus, the impactor successively slows down and finally sticks to the cluster.

\begin{figure}[t]
    \begin{center}
        \includegraphics[width=8 cm]{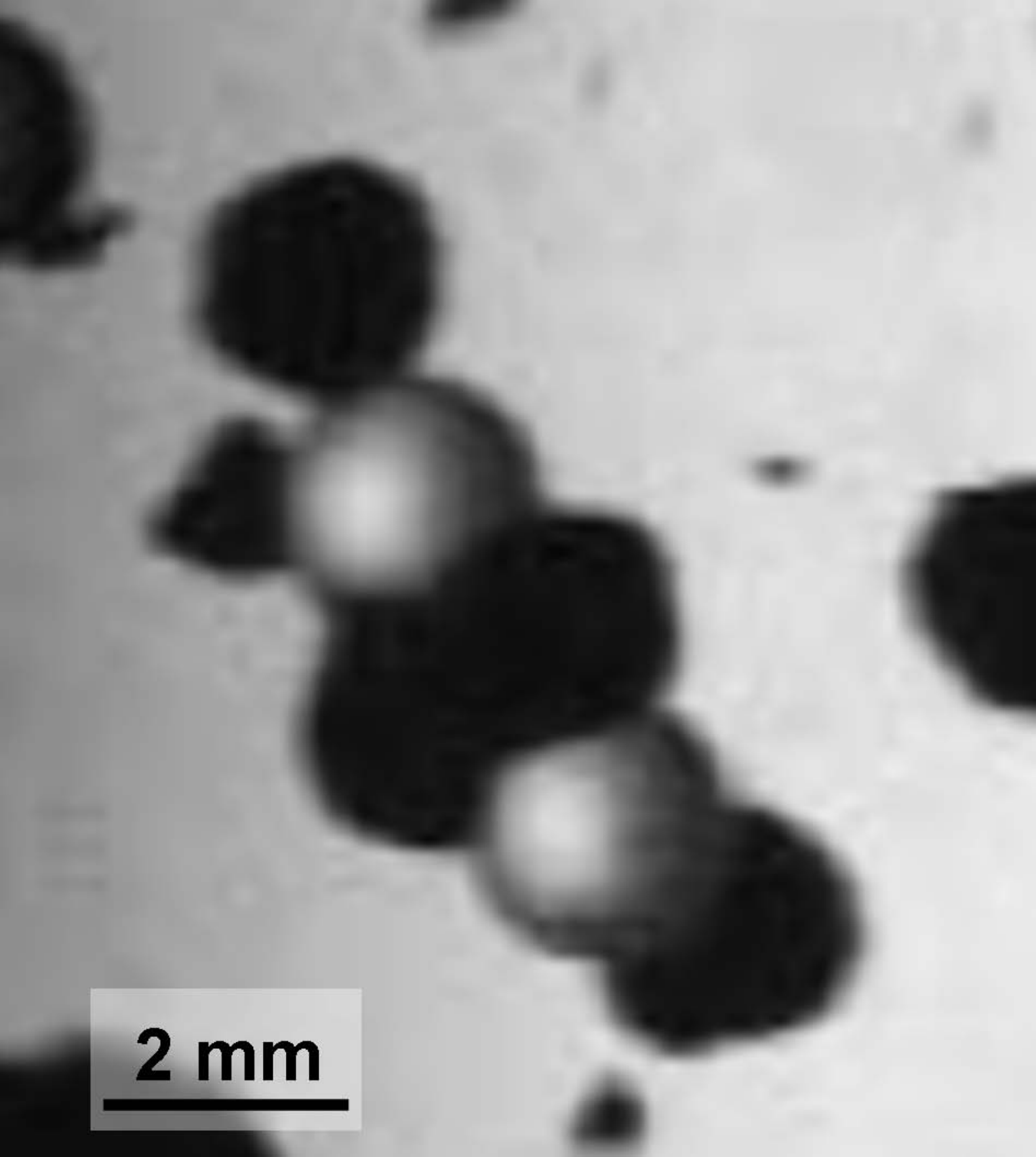}
        \caption{\label{fig:cluster_image}Example of a cluster that formed in the experiments of un-coated glass beads (2mm diameter, bright spherical objects in the image) and dust aggregates of the same size (black objects in the image). It is obvious that both particle types are alternatingly arranged.}
    \end{center}
\end{figure}

The formation of larger clusters by capturing free particles can also been found in the experiments with glass beads and dust aggregates, with the restriction that sticking collisions can only occur between different species and not among the same ones. This effect leads to an alternating assembly between the dust aggregates and the glass beads, which can be qualitatively seen in Fig. \ref{fig:cluster_image}.

In the experiment shown in Fig. \ref{fig:cluster_sequence} the initial cluster consisted of 15-20 dust-coated chondrule analogs and possessed a size of $< 1$ cm. At the end of the experiment, this cluster had grown to a size of $\sim 2$ cm (probably limited by the diameter of the experiment tube) and consisted of 60-70 chondrule analogs. Typical impact velocities into this cluster were a few $\rm cm \, s^{-1}$. Thus, we could show with our experiments that the growth of several-cm-sized bodies under solar-nebula conditions is possible for dust-coated chondrules. Therefore, chondrules could be a key to our understanding of the formation of planetesimals in the protoplanetary disk.

\section{Concluding Remarks}

We observed sticking in free collisions between dust-coated glass beads of 2 and 3 mm diameter as well as between dust aggregates and un-coated glass beads at velocities of up to $\sim$8 \cms. This is an unexpected result, in as far as neither solid particles of the same size \citep[see, e.g.,][]{FoersterEtal:1994, LorenzEtal:1997}, nor porous 1.5 mm  dust aggregates (see Paper I) stick to each other at those velocities. We found that solid particles with either porous dust rims or in a system with sufficiently many dust aggregates apparently show a considerably enhanced sticking behavior. This is obviously caused by a favorable combination of high inertia (carried by the solid glass beads) and deformability (carried by either the dust rims or the dust aggregates), which increases both the dissipation of kinetic energy and the contact area between the collision partners, thus enabling sticking at unexpectedly high velocities.

In Paper I, we adapted a model of \citet{ThorntonNing:1998} to deduce the sticking velocity of porous aggregates and the agreement was satisfactory. However, extrapolating these results by taking their Equation (54), namely $v_{\rm stick} \propto r^{-5/6} \, \rho^{-1/2}$, thus assuming that the effective surface energy and Young's modulus of the dust rim were identical to the values assumed in Paper I, we get, due to the larger particle radii ($r=1$ mm instead of $r=0.5$ mm in Paper I) and mass densities ($\rho = 2500 \rm \, kg\, m^{-3}$ instead of $\rho = 700 \rm \, kg\, m^{-3}$ in Paper I), a prediction for the sticking threshold a factor 3.4 {\it smaller} than that for the dust aggregates in Paper I. Taking a value of $v_{\rm stick} = 1.8 \rm \, mm\, s^{-1}$ (within an uncertainty of a factor of two) for the 50\% sticking probability for the dust aggregates in Paper I, we then arrive for our dust-coated glass beads at a prediction of $v_{\rm stick} = 0.5 \rm \, mm\, s^{-1}$, which is obviously way too low compared to our findings in this paper that this value should be around $v_{\rm stick} \approx 10 \rm \, mm\, s^{-1}$ (see Figs. \ref{fig:sicking} and \ref{fig:sicking_beadsanddust}). Thus, we can conclude that a model describing the collision and sticking behavior of dust-rimmed solid particles or that of solid particles with dust aggregates is still lacking. To describe those particles a model that takes plastic deformation during the penetration phase into account is required and this was not done by \citet{ThorntonNing:1998}.

The velocity range for sticking observed in our experiments is highly relevant for millimeter-sized particles. As \cite{WeidlingEtal:2009} showed in their Fig. 8, the relative velocities for particles of 1 cm radius are below 1 \cms\ if they assume a high solar-nebula-surface density considered by \citet{Desch:2007} and of the order of 1 \cms\ for a minimum-mass solar nebula and mm-sized particles. Thus, a growth to somewhat larger clusters into the cm-size region seems plausible as was already shown in Sect. \ref{sec:cluster_formation}. In subsequent collisions of centimeter-sized clusters, new ways for the dissipation of energy as qualitatively shown in Sect. \ref{sec:cluster_formation} may lead to further growth. Higher dust-to-gas ratios as expected in chondrule forming regions \citep{AlexanderEtal:2008} and from numerical simulations \citep{JohansenEtal:2007, CuzziEtal:2008} may have an additional positive effect on the growth by further decreasing the relative velocities.

Finally, our experiments indicate a possible way for the formation of the parent bodies of chondrites. It still needs to be solved how stable the observed clusters are against disruption and how the results depend on material parameters, i.e., dust-grain size, dust composition, and possibly thermal history. This will be done in future experiments.

\subsection*{Acknowledgements}
E.B. and C.G. thank the Deutsche Forschungsgemeinschaft (DFG) for support under grants Bl 298/13-1 and Bl 298/7-2. We thank the Deutsches Zentrum f\"ur Luft- und Raumfahrt (DLR) for support in the development of the MEDEA hardware (grant 50WM0936) and for providing us with drop tower flights. We also thank Torsten Krumscheid for aiding us in the image analysis of the drop tower experiments.

\bibliographystyle{aa}
\bibliography{literatur}

\end{document}